\documentclass[letter]{aa}  

\usepackage{graphicx}
\usepackage{txfonts}
\newcommand{\Msun}{\rm{M}_\odot}
\newcommand{\Rsun}{\rm{R}_\odot}
\newcommand{\Rflat}{R_{\rm{flat}}}
\newcommand{\Rhalt}{R_{\rm{halt}}}
\begin{document}

   \title{The final orbital separation in common envelope evolution}


   \author{M. Politano
          \inst{1}
          }

   \institute{Marquette University, Department of Physics, P.O. Box 1881, Milwaukee, Wisconsin, 53201-1881, U.S.A.\\
              \email{michael.politano@marquette.edu}
             }

   \date{Received January 25, 2021; }

 
  \abstract
 {In the majority of population synthesis calculations of close binary stars, the common envelope (CE) 
phase is modeled using a standard prescription based upon the conservation of energy, known as the alpha prescription.  In this prescription, the orbital separation of the secondary and giant core at the end of the CE phase is taken to be the orbital separation when the envelope becomes unbound.  However, recent observations of planetary nebulae with binary cores (BPNe), believed to be the immediate products of CE evolution,  indicate orbital periods that are significantly shorter than predicted by population synthesis models using the alpha prescription.  We argue that unbinding the envelope provides a necessary, but not sufficient, condition to escape a merger during CE evolution.  The spiral-in of the secondary must also be halted. This requires the additional dynamical constraint that the frictional torque on the secondary be reduced to approximately zero.  In this paper, we undertake a preliminary examination of the effect of adding this dynamical constraint in population synthesis calculations of BPNe.  We assume that the frictional torque will be sufficiently reduced when the secondary enters a region within the giant where the mass-radius profile is flat.  We crudely estimate the location of this region as a function of the core mass based upon existing stellar models of AGB stars between 1 and 7 $\Msun$. We calculate a theoretical orbital period distribution of BPNe using a population synthesis code that incorporates this dynamical constraint along with the alpha prescription.}


   \maketitle
%

\section{Introduction}

The common envelope (CE) phase is believed to play a key role in the evolution of many close binary stars, yet a firm theoretical understanding of this phase remains elusive.  In a typical CE scenario, the initially more massive star (the primary) is a giant and the less massive star (the secondary) is a low-mass main sequence star. Either as a result of tidal capture or unstable mass transfer due to Roche lobe overflow, the secondary star becomes engulfed in the envelope of the primary.  The orbit of the secondary decays due to frictional drag and the secondary spirals in toward the core of the primary.  There are two possible outcomes of such an evolution: Either a merger occurs or sufficient orbital energy is released on a short enough timescale to eject the envelope, leaving the embedded components intact but in a much closer orbit. 

Detailed hydrodynamical calculations of CE evolution have been made by several researchers using both grid-based and SPH codes (e.g., \citealp{iva13}; \citealp{iac19}).  In many of the grid-based calculations, the final outcome of the evolution was not able to be determined, primarily because of the large difference in spatial scales that need to be resolved (e.g., \citealp{san98}).  In some cases where the evolution was followed to completion, either the envelope was not able to be ejected or the final orbital separations were much larger than observed in post-CE systems, or both (e.g., \citealp{ric12,pas12,ohl16,san20}).  \citet{nan16} were able to unbind the envelope using recombination energy.   Nevertheless, \citet{iac19}, who conducted a detailed comparison of all 3-D simulations of the CE phase, conclude that the final orbital separations in all simulations must be considered upper limits due to a lack of spatial resolution or the size of the softening length applied to the gravitational potential of the particles. Further, the parameter space of post-CE binaries is not well covered by the simulations. Until these limitations are overcome, hydrodynamical calculations of the CE phase will not provide the necessary input for modeling populations of close binary stars using population synthesis techniques.

In order to make progress, a very simple model of the CE phase was proposed based upon the conservation of energy \citep{tut79}.  In this model, the envelope is assumed to be ejected (somehow) if sufficient orbital energy is released during the spiral-in to unbind it before a merger occurs.  This so-called alpha prescription, shown in Equation~\ref{alpha}, was developed to estimate the orbital separation at which the envelope would become unbound.
\begin{equation}\label{alpha}
 -\alpha \left(\frac{Gm_cm_s}{2a_{\rm ej}} - \frac{Gm_pm_s}{2a_i}\right) = E_{\rm{bind}.}
.\end{equation}
In this prescription, $m_c$, $m_p$, $E_{\rm{bind}}$, and $a_i$ are the core mass of the primary, the total mass of the primary, the binding energy of the primary's envelope, and the orbital separation, respectively, all at the onset of the CE phase, $m_s$ is the mass of the secondary, $a_{\rm ej}$ is the orbital separation when the envelope becomes unbound, and $\alpha$ is the so-called CE efficiency parameter, which is an input parameter that represents the efficiency at which orbital energy is transferred to the envelope.  Effectively, the many uncertainties in CE evolution have been encapsulated in the efficiency parameter, whose value is unknown.  It is important to note that the alpha prescription assumes that any remaining envelope material interior to the orbit (after ejection) is small enough to have a negligible effect on the final orbital separation.  Thus, in the alpha prescription, $a_{\rm ej}$ in Equation~\ref{alpha} is considered to be the orbital separation at the end of the CE phase, which we shall denote as $a_{\rm f}$.

\begin{figure}
\includegraphics[width=\columnwidth]{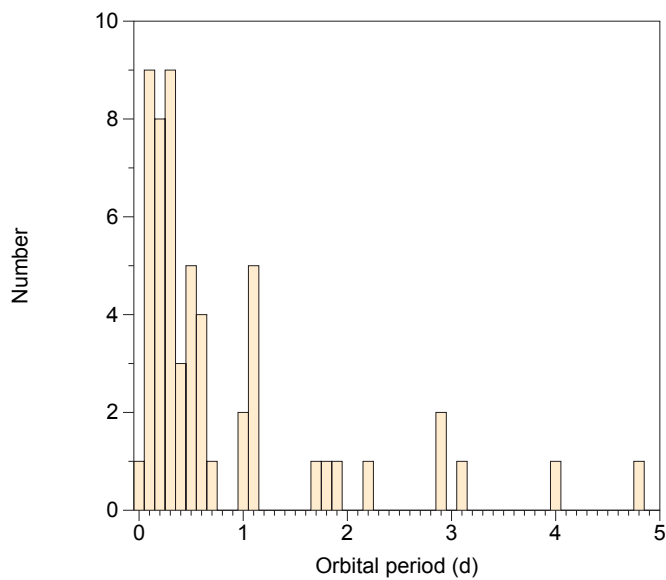}
\caption{Observed distribution of orbital periods in BPNe.}\label{obs_porb}
\end{figure}

Observations of post-CE binaries (PCEBs) have increased dramatically in the last ten years.  Of particular importance are observations of planetary nebulae with binary cores (BPNe) since these systems are believed to have just exited the CE phase.  
Currently, there are $\sim$60 BPNe with known orbital periods\footnote{See online catalog maintained by David Jones at drdjones.net/bcspn.}.  A distribution of these periods
is shown in Figure~\ref{obs_porb}.  This distribution is sharply peaked near P$_{\rm{orb}}$ = 1/3 day and contains few systems with orbital periods greater than 1 day. 

Model populations of BPNe have been calculated independently by several researchers using population synthesis codes that employ the alpha prescription \citep{deK90,yun93,han95,mdm06}.  For common input parameters ($\alpha = 1$ and a flat initial mass ratio distribution), the orbital period distributions in these model populations are quite similar, but they do not match the observed distribution.   Instead, the model orbital period distributions are broadly peaked, contain a large number of systems with periods greater than 1 day, and have a long-period tail. 

The cause of the discrepancy between the theoretical period distributions and the observed orbital period distribution in BPNe remains unknown.   Independent analyses conclude that the dearth of BPNe with orbital periods longer than 1 day is not likely a result of observational biases \citep{dhs08,rm08,mis09,dav10,jb17}.  However, observational selection effects cannot be completely ruled out. 

In this paper, we examine the effect of adding a dynamical constraint to determining the final orbital separation. Namely, we require that, in addition to unbinding the envelope, the frictional drag on the secondary must also be severely reduced in order to halt the spiral-in and avoid a merger.  This dynamical constraint was proposed by \citet{ybt95} and \citet{tt96}, but it has never been incorporated into population synthesis calculations.  Using data presented in \citet{tt96}, we have developed a very crude relationship between the orbital separation, at which spiral-in is halted, and the core mass of the giant primary.  We have incorporated this relationship into our population synthesis code and calculated new theoretical orbital period distributions for BPNe.  We find that these new distributions are in better agreement with the observed period distribution in BPNe.

In section 2, we discuss the constraint in more detail and its incorporation into our code.  In section 3, we present the theoretical orbital period distributions in our model BPNe populations calculated with the dynamical constraint. We discuss these results and present our conclusions in section 4.

\section{Method}

\begin{figure}
\includegraphics[width=\columnwidth]{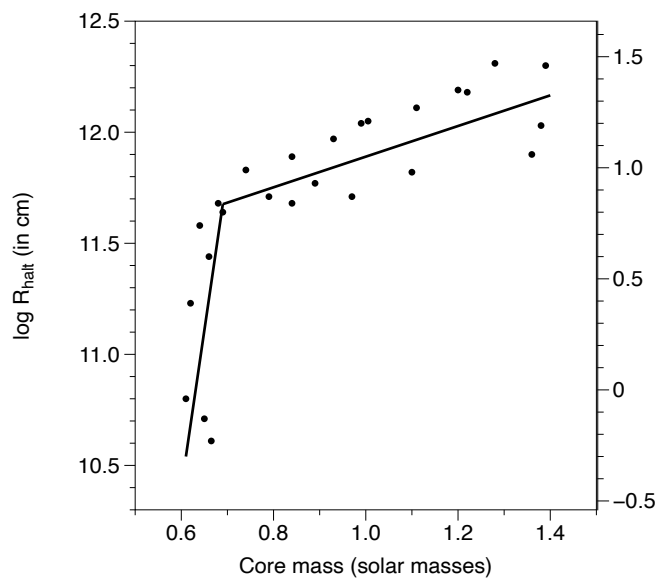}
\caption{Radius at which the spiral is halted versus the core mass of the giant at various evolutionary stages and for various initial masses. We note that $\Rhalt$ = $\Rflat$/6.  The data points are taken from Figure 8 in \citet{tt96} and the solid line is our linear fit to their data. The right scale on the y-axis is log $\Rhalt$ in $\Rsun$.}\label{Rhalt}
\end{figure}

\citet{ybt95} and \citet{tt96} argue that while the release of enough orbital energy to unbind the envelope is a necessary condition to avoid a merger, it is not a sufficient condition. The frictional torque on the secondary must also become sufficiently small, otherwise the orbital decay will not be halted and the secondary will continue to spiral in toward the core.  The authors maintain that in order for this frictional torque to be reduced significantly, the envelope must contain a region where the mass-radius profile is flat (see Figure 6 in \citealp{ybt95}).    \citet{tt96} estimated the outermost radial coordinate at which this flat region begins using the condition that 
$V=-d\,$\rm{ln}$\,p/d\,$\rm{ln}$\,r$ is a minimum (see e.g., \citealp{sch58}).  In this expression, $p$ is the pressure and $r$ is the radial coordinate.  Essentially, since $V$ is proportional to $M(r)/r$, where $M(r)$ is the mass contained within radius $r$, $V$ will decrease outward in a region where the mass-radius profile is flat until $M(r)$ begins to increase significantly.  We denote this outermost radial coordinate as $\Rflat$. 

 \citet{ybt95} and \citet{tt96} note that the frictional torque will not immediately drop to zero at $\Rflat$, since there is still sufficient material within the gravitational interaction region of the secondary.  They estimated from their hydrodynamical calculations that the frictional torque will be reduced sufficiently to halt spiral-in at a radial coordinate between approximately $\Rflat$/10 to $\Rflat$/3, and they chose 
$\Rflat$/6 as an estimate.    Using stellar models of giant stars between 1 and 7 $\Msun$, \citet{tt96} determined $\Rflat$ for giants at various points in their evolution.  Choosing $\Rflat$/6 as the radius at which the spiral-in is halted, they plotted this critical radius versus the core mass of the giant at various evolutionary stages and for various initial primary masses (Figure 8 in \citealp{tt96}).

We show in Figure~\ref{Rhalt} a similar plot in which the data points for AGB stars are taken from Figure 8 in \citet{tt96} and the lines represent our crude linear fits to their data, which are expressed mathematically as follows:  
\begin{equation}\label{linearfits}
    \log{(\Rflat/6)} = 
\begin{cases}
    9.0\,m_c + 5.4                       & \text{if } 0.55 \leq m_c \leq 0.70\\
    0.7857\,m_c + 11.15             & \text{if } 0.70 \leq m_c \leq 1.4,\\
\end{cases}
\end{equation}
where $\Rflat$ is in centimeters and $m_c$ is in $\Msun$. We have neglected RGB stars in Figure~\ref{Rhalt}, since the data in Figure 8 of \citet{tt96} are much scarcer for RGB stars compared to AGB stars.  


\citet{ybt95} and \citet{tt96} further maintained that the envelope must be ejected before the orbital deceleration is significantly reduced.  Once the orbital decay timescale increases significantly, even if enough orbital energy is available to unbind the envelope, it would be released on too long of a timescale to be effectively converted into mechanical energy and drive mass motion.  Instead, it would be converted into thermal energy and radiated away, ultimately leading to a merger.

The population synthesis code used to calculate the model populations of BPNe has been described extensively in the literature (e.g., \citealp{pol10}).  Since the only modification made to the code is the prescription for the CE phase, we refer the reader to the above paper for detailed information about the code. Here we focus on the incorporation of the dynamical constraint and we describe only the most salient features of the code.  

In modeling the CE phase, we still used equation~\ref{alpha} to determine $a_{\rm ej}$, which is the orbital separation at which the CE becomes unbound. The envelope binding energy was taken directly from the stellar models used in the population synthesis. We next determined $\Rflat$ from Equation~\ref{linearfits}.    The radius at which the spiral-in is halted is denoted by $\Rhalt$.  In our standard model, we assumed $\alpha$ = 1 and $\Rhalt$ = $\Rflat$/6 (following \citealp{tt96}), but we investigated other choices for these parameters.  Finally, we determined the Roche lobe radius of the secondary at an orbital separation equal to $\Rhalt$ and compared it to the radius of the secondary, $R_s$. In order for the system to avoid a merger, we required that two criteria be met:  (1) The envelope must have become unbound before the torque on the secondary was severely reduced and the orbital decay timescale became too long to drive ejection (i.e., $a_{\rm ej} >$ $\Rhalt$); and (2) the Roche lobe radius of the secondary was greater than the radius of the secondary when the orbital separation equaled $\Rhalt$ (i.e., $R_{\rm L,s, halt} > R_s$).  These criteria are nicely summarized by the following inequality,
\begin{equation}\label{inequality}
R_s < R_{\rm L,s, halt} <  \Rhalt < a_{\rm ej}.
\end{equation}
We note that $R_{\rm L,s, halt} = r_{\rm L}\,\Rhalt$, where $r_{\rm L}$ is the dimensionless Roche lobe radius taken from \citet{egg83}.

If the inequality in equation~\ref{inequality} cannot be satisfied, a merger is assumed and the system is removed from the population.  If the inequality can be satisfied, we assume that no further decay of the orbit occurs once the secondary reaches $\Rhalt$ and we set  $a_{\rm f}$ = $\Rhalt$.    

In our calculations, we began with 10$^9$ zero-age main sequence binaries. We assumed a Miller-Scalo initial mass function \citep{ms86} for the primary masses, an orbital period distribution that is flat in log $P$ \citep{abt83}, and an initial mass ratio distribution that is flat, $g(q)\,dq = 1\,dq$ \citep{dm91}.  We assumed a Galactic age of 10$^{10}$ yrs, representative of a disk population, and assumed a constant star formation rate.  We used full stellar models to follow the evolution of both stars and to determine the binding energy of the envelope at the onset of the CE phase.  We only considered primaries on the AGB at the onset of the CE phase.  Lastly, we assumed that the radius of the secondary remains unchanged during CE evolution.

\section{Results}

\begin{figure}
\includegraphics[width=\columnwidth]{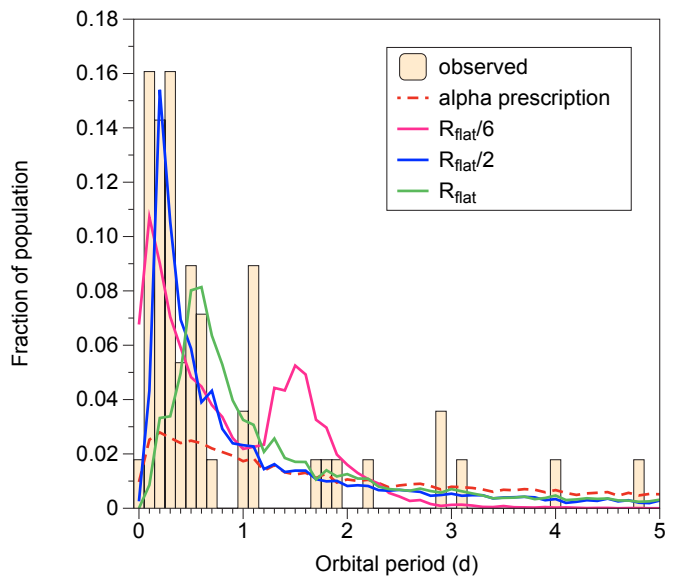}
\caption{Model BPNe orbital period distributions for various values of $\Rhalt$.  The observed orbital period distribution in BPNe (histogram) and the orbital period distribution using the standard alpha prescription (red dashed line) are also shown for comparison.   All distributions have been normalized to unity for the sake of comparison.}\label{Rhalt_compare}
\end{figure}

\begin{figure}
\includegraphics[width=\columnwidth]{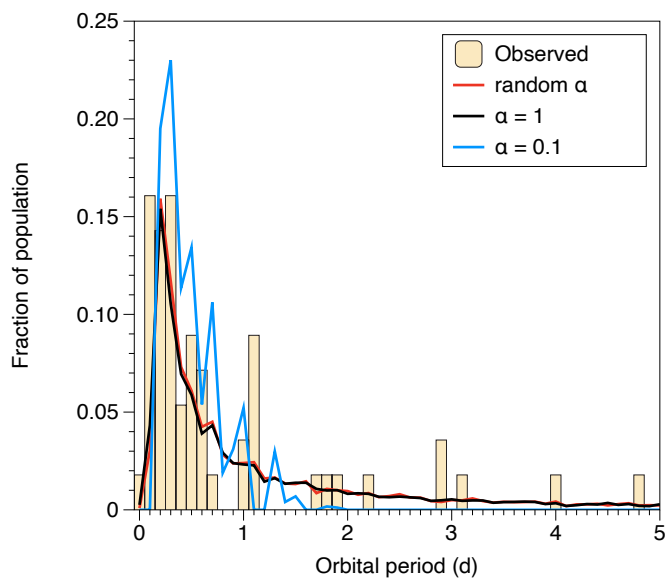}
\caption{Model BPNe orbital period distributions for various choices of $\alpha$.  In all cases, $\Rhalt = \Rflat/2$.   All distributions have been normalized to unity for the sake of comparison.}\label{alpha_compare}
\end{figure}

In Figure~\ref{Rhalt_compare}, we show the BPNe orbital period distributions calculated using the CE prescription with the dynamical constraint for various values of $\Rhalt$, ranging from $\Rflat$/6 to $\Rflat$, assuming $\alpha = 1$.  In Figure~\ref{alpha_compare}, we show the model period distributions for various choices of $\alpha$, assuming $\Rhalt = \Rflat/2$.  

We find that the inclusion of the dynamical constraint as described in section 2 has the following effects.

First, for all choices of $\Rhalt$ and $\alpha$, the fraction of systems with orbital periods less than 1 day is significantly larger than the fraction found using the standard alpha prescription.

Second, for all choices of $\Rhalt$ and $\alpha$, the distribution is peaked, in contrast to the broad distribution found using the standard alpha prescription. We find that the location and height of the peak in the $\Rhalt = \Rflat/2$ distribution best matches the observed distribution. 

Third, the shape of the distribution is not very sensitive to the choice of $\alpha$.  Indeed, choosing a random value for $\alpha$ between 0 and 1 yields a nearly identical distribution to the distribution where alpha = 1.  Only when $\alpha$ is globally very small is there any notable difference.

\section{Discussion and conclusions}

The alpha prescription was originally proposed as a way to crudely estimate the orbital separation at the end of the CE phase in lieu of a detailed model of CE evolution.  However, it is unlikely that the many uncertainties in CE evolution can be distilled into a single global parameter whose value can be constrained by observations. Attempts to do so for PCEBs have succeeded only for very specialized choices of input parameters, for example, $\alpha \la 0.2-0.3$ \citep{zor10}. 

One of the difficulties in modeling the CE phase is the wide range of timescales involved in the evolution.  The rapid spiral-in phase occurs on a dynamical timescale and can only be properly modeled using a 3-D hydrodynamical code, while the slow spiral-in phase (also called the self-regulated phase) occurs on a thermal timescale and requires a 1-D stellar evolution code, which contains more detailed physics. In their comprehensive review of CE evolution, \citet{iva13} cite understanding the transition between the rapid and slow spiral-in phases as a high priority in advancing our understanding of CE evolution.  An important step in that process is determining the orbital separation at which the orbital decay slows down.   Once the secondary enters the flat mass-radius region of the envelope, the orbital decay time will begin to rise considerably (e.g., \citealp{tt96}).  We suggest that the transition between the two phases occurs somewhere within this region. 
 
The improved agreement between the theoretical and observational orbital period distribution of BPNe suggests that inclusion of a dynamical constraint in population synthesis calculations may be an important consideration.  However, the analysis we present here is very crude.  In particular, a more detailed analysis of the relationship between $\Rflat$, the core mass of the giant, and the total mass of the giant is needed (e.g., \citealp{lop15}).  We suggest a grid of $\Rflat$ for giants of various total mass and core mass needs to be calculated from detailed stellar models of both RGB and AGB stars.  
The question of whether there are a significant number of BPNe with RGB progenitors remains open (e.g., \citealp{hil17,jon20}).  Nevertheless, they should be included in the population synthesis (e.g., \citealp{hal13}).  Secondly, the frictional torque per unit volume, $\tau$, is a function of the density, $\tau = f \rho {V_{\rm rel}}^2/2 \pi$, where $\rho$ is the local density,  $V_{\rm rel}$ is the velocity of the secondary relative to the envelope, and $f$ is a parameter related to the Mach number \citep{ybt95}.  Once $\Rflat$ is reached, there is uncertainty as to the orbital separation at which the frictional torque becomes effectively zero. \citet{tt96} assumed $\Rhalt = \Rflat/6$, but further investigation is needed to determine a more precise relationship.

With the inclusion of the dynamical constraint, the final orbital separation primarily depends upon the evolutionary state of the giant star and is relatively independent of the secondary mass (as long as the envelope can become unbound). If the dependence on the total mass of the giant is weak, there should be a correlation between the distribution of core masses and the distribution of orbital periods in BPNe. Currently, there is an insufficient number of BPNe in which both the orbital period and the white dwarf mass are known.  However, we expect that continued observations of BPNe will eventually allow this prediction to be tested.

\begin{acknowledgements}
     We gratefully acknowledge very useful conversations with A. Ruiter, T. Hillwig, and D. Jones. We also thank the anonymous referee for very helpful comments.
\end{acknowledgements}

%
%

\end{document}